\documentclass[aip,cha,amsmath,amssymb,reprint]{revtex4-2}

\usepackage{graphicx}
\usepackage{dcolumn}
\usepackage{bm}
\usepackage{subfigure}
\usepackage{multirow}
\usepackage[utf8]{inputenc}
\usepackage[T1]{fontenc}
\usepackage{mathptmx}
\usepackage{etoolbox}
\usepackage{lipsum}
\usepackage{xcolor}
\usepackage{xspace, units}
\usepackage{soul}

\newcommand{\kthresh}{\ensuremath{k_{\mathrm{th}}}}

\newcommand{\algcon}{\ensuremath{\lambda_{2}}}
\newcommand{\rewprob}{\ensuremath{p_{\mathrm{r}}}}
\newcommand{\gameps}{\ensuremath{\gamma_{\epsilon}}}
\newcommand{\gamlam}{\ensuremath{\gamma_{\lambda_{2}}}}

\begin{document}

\author{Christian Hechler}%
\email{christian.hechler@uni-bonn.de}
\affiliation{Department of Epileptology, University of Bonn Medical Centre, Venusberg Campus 1, 53105 Bonn, Germany}
\affiliation{Helmholtz Institute for Radiation and Nuclear Physics, University of Bonn, Nussallee 14–16, 53115 Bonn, Germany}
\author{Timo Br\"ohl}
\email{timo.broehl@ukbonn.de}
\affiliation{Department of Epileptology, University of Bonn Medical Centre, Venusberg Campus 1, 53105 Bonn, Germany}
\affiliation{Helmholtz Institute for Radiation and Nuclear Physics, University of Bonn, Nussallee 14–16, 53115 Bonn, Germany}
\author{Ulrike Feudel}
\email{ulrike.feudel@uni-oldenburg.de}
\affiliation{Theoretische Physik/Komplexe Systeme, ICBM, Carl von Ossietzky Universität Oldenburg, Carl-von-Ossietzky-Straße 9-11, 26111 Oldenburg, Germany}
\author{Klaus Lehnertz}
\email{klaus.lehnertz@ukbonn.de}
\affiliation{Department of Epileptology, University of Bonn Medical Centre, Venusberg Campus 1, 53105 Bonn, Germany}
\affiliation{Helmholtz Institute for Radiation and Nuclear Physics, University of Bonn, Nussallee 14–16, 53115 Bonn, Germany}
\affiliation{Interdisciplinary Center for Complex Systems, University of Bonn, Brühler Straße 7, 53175 Bonn, Germany}

\title{Complex network topological and spectral determinants of extreme events} 

\makeatletter
\def\@email#1#2{%
 \endgroup
 \patchcmd{\titleblock@produce}
  {\frontmatter@RRAPformat}
  {\frontmatter@RRAPformat{\produce@RRAP{*#1\href{mailto:#2}{#2}}}\frontmatter@RRAPformat}
  {}
}%
\makeatother

\date{\today}

\begin{abstract}
We study the impact of the coupling topology on the ability of various networked dynamical systems to generate extreme events.
By determining the coupling strength that is necessary to generate an extreme event in the collective dynamics of a given system, we observe a power-law-like relationship between this coupling threshold and both topological (edge density) and spectral (algebraic connectivity) properties of various coupling topologies.
Interestingly, this relationship appears to be largely independent of both the investigated system and the underlying mechanism to generate extreme events.
This may indicate that the observed relationship is primarily mediated by aspects of the coupling topology. 
\end{abstract}


\maketitle 

\begin{quotation}
When investigating the emergence of extreme events in high-dimensional networked dynamical systems, one often faces the problem of carefully selecting control parameter settings for both node dynamics and coupling topology.
Lacking detailed knowledge about the impact of structural and functional aspects of such systems on the generation of extreme events, finding appropriate control parameter settings is often cumbersome.
We performed a brute-force search for the threshold of a crucial control parameter, namely the coupling strength, at which extreme events emerge in the dynamics of a variety of networked systems capable of intrinsically generating extreme events due to diverse underlying dynamical mechanisms.
Surprisingly, we observed that the coupling threshold can easily be derived from power-law-like relationships with easy-to-identify topological and spectral properties of various coupling topologies.
The relationships not only provide useful guidance for future model studies to quickly identify control parameter settings, but also highlight the importance of considering topological and spectral network properties to further improve our understanding of generation and mitigation of extreme events.
\end{quotation}
\frenchspacing
\section{Introduction}
\label{sec:intro}
Extreme climate and weather conditions~\cite{Fischer2025}, large-scale disruptions in supply networks (e.g. power outages)~\cite{dobson2007Chaos}, market crashes~\cite{Feigenbaum2001}, rogue waves in the ocean~\cite{Dysthe2008} or in optical systems~\cite{solli2007Nature}, harmful algal blooms in marine ecosystems~\cite{Dai2023}, or epileptic seizures in the human brain~\cite{Lehnertz2023}~--~these are just a few examples of serious and severe phenomena occurring in natural or man-made systems. 
Such rare and significant extreme events~\cite{Albeverio2006} have a lasting impact on system constituents or even the system as a whole.
Given the devastating consequences entailed by extreme events, huge efforts are made to minimize their impact, mitigate their consequences, or prevent them as well as to identify early warning signs. 
However, forecasting extreme events remains a major challenge given the complexity of natural systems. 

Even greater is the scientific interest in understanding extreme events and their underlying mechanisms~\cite{Ghil2011,chowdhury2022PhysRep,Alvre2024}.
For many real-world systems, active experiments on extreme events are technically infeasible or ethically unacceptable, as even small perturbations may trigger such an event and cause significant damage. 
Nevertheless, important insights into dynamical mechanisms underlying the generation of extreme events can be gained from investigating the dynamics of model systems capable of undergoing critical transitions.
The latter may be related to bifurcations~\cite{Horsthemke1984,Kuehn2011,Ashwin2012,Feudel2023}, or to other mechanisms that have been less studied to date such as boundary crisis~\cite{Grebogi1983,Osinga2000}, attractor bubbling~\cite{Ashwin1994,Venkataramani1996a,Venkataramani1996b}, 
intermittency~\cite{Saha2017,Mishra2020}, blowout bifurcations~\cite{Ott1994,Zhang2020}, saddle escape~\cite{Kuehn2015}, explosive synchronization~\cite{Kuehn2021}, or oscillation death~\cite{Zou2021,Koseska2013}.
Given the intricate interplay between structure and function in networked dynamical systems~\cite{Boccaletti2006,Boccaletti2014,Boccaletti2016}, understanding the emergence of extreme events --~as a collective phenomenon~-- in systems with complex coupling topologies has become an active and interdisciplinary field of research~\cite{Kishore2011,ansmann2013PhysRevE,karnatak2014PhysRevE,Ansmann2016,Chowdhury2019,Broehl2020,Malik2020,Ray2020,Ray2022,Roy2024,Broehl2025,CubillosCornejo2025,Mehrabbeik2025,Ramasamy2025,Shashangan2025}.

Although various mechanisms leading to extreme events in coupled dynamical systems have been identified~\cite{chowdhury2022PhysRep}, the specific influence of properties of the coupling topology on these phenomena is only barely investigated. 
This may be due to the fact that the occurrence of extreme events in complex networks is not only guided by control parameter settings but also by the size and structure of the network~\cite{karnatak2014PhysRevE}, which considerably complicates the derivation of robust estimates in the large-scale limit.
Addressing this issue, we here explore the impact of the coupling topology on the ability of various networked dynamical systems to generate extreme events, as well as the conditions under which they occur. 

\section{Methods}
We investigate the dynamics of networked systems of the form
\begin{equation}
    \dot{x}_i = f(x_i)+\frac{k}{N-1} \sum_{j=1}^N A_{ij} H(x_i,x_j),
		\label{eq:netsys}
\end{equation}
where $f(x_i)$ describes the eigendynamics of subsystem $i \; (i\in\left\{1,\ldots,N\right\})$. 
In particular, we regard networks of coupled FitzHugh-Nagumo oscillators~\cite{ansmann2013PhysRevE,karnatak2014PhysRevE} (system S1), of coupled memristive Hindmarsh–Rose neurons~\cite{vijay2024EurPhysJ} (system S2), of coupled periodically forced Li\'enard-type oscillators~\cite{kingston2017PhysRevE} (system S3), and of coupled R\"ossler oscillators~\cite{tirabassi2025Arxiv} (system S4) (for details, see Appendix~\ref{app:systems}). 
While the generation of extreme events in system S4 is related to a bubbling transition that gives rise to sudden desynchronization bursts~\cite{Ashwin1994}, the generating mechanism in systems S1 and S2 is related to an interior crisis~\cite{Grebogi1987}. 
In system S3, extreme events can occur either via an interior crisis or via intermittency~\cite{kingston2017PhysRevE}. 

In Eq.~\ref{eq:netsys}, the coupling strength is denoted by $k$ and $H$ is the coupling function. 
The symmetric adjacency matrix ${\bf A} \in \{0,1\}^{N \times N}$ has entries $A_{ij} = A_{ji} = 1$, if and only if subsystems $i$ and $j$ are coupled. 
We regard coupling topologies that are based on paradigmatic network models, namely, random networks\cite{erdos1959PublMathDebr}, small-world networks~\cite{watts1998Nature} (with different rewiring probabilities $\rewprob$), and scale-free networks\cite{albert2002RevModPhys}, each of which connects $N$ vertices with different number of edges.
Control parameters for network generation are chosen to allow for a comparable edge density $\epsilon ={2E}{(N(N-1))^{-1}}$ among the coupling topologies, while retaining the respective defining topological properties.
$E$ denotes the overall number of edges.
Networks are connected and do not possess unconnected subnetworks.

We are interested in elucidating whether there exists a dynamics- and coupling-topology-dependent threshold $\kthresh$ for the coupling strength that leads to the emergence of extreme events in a system’s dynamics.
Since there is so far no commonly accepted recipe to determine $\kthresh$, we here perform a brute-force search:
for a given networked system (of size $N$) with adjacency matrix ${\bf A}$ and in case of small-wold coupling topology, a given rewiring probability $\rewprob$, we adaptively determine the minimum coupling strength required for the emergence of extreme events in the dynamics of systems S1 -- S3, and the maximum coupling strength at which extreme events can still be observed in system S4, since attractor bubbling in this system arises as the coupling strength is decreased.

We conjecture that topological and spectral properties of the network structure underlying the coupling play a decisive role for the generation of extreme events.
As an indicator for the former, we consider the edge density $\epsilon$, which discloses vital aspects of network connectedness. 
Increasing the edge density improves inter-vertex reachability (due to a decreased network diameter) and can trigger phase transitions and synchronization~\cite{ricci2012,squartini2013,rakshit2020}.
We have $0.04 \leq \epsilon \leq 0.99$ for random and small-world coupling topologies, and $0.04 \leq \epsilon \leq 0.5$ for scale-free coupling topologies.
For the latter network property, we consider the algebraic connectivity~\cite{fiedler1973} $\algcon$, which is the second smallest eigenvalue of the corresponding Laplacian matrix ${\bf L}$, whose elements are $L_{ij} = \kappa_i\delta_{ij} - A_{ij}$, where $\delta_{ij}$ is the Kronecker delta, and $\kappa_i$ denotes the degree of the $i$th vertex. $\algcon$ signifies  a network's intrinsic structural stability and robustness~\cite{oehlers2021}, and it indicates how quickly a connected network ($\algcon>0$~\cite{atay2005}) can diffuse information and synchronize~\cite{arenas2008}.
\begin{figure}[htbp]
   \centering   \includegraphics[width=\linewidth]{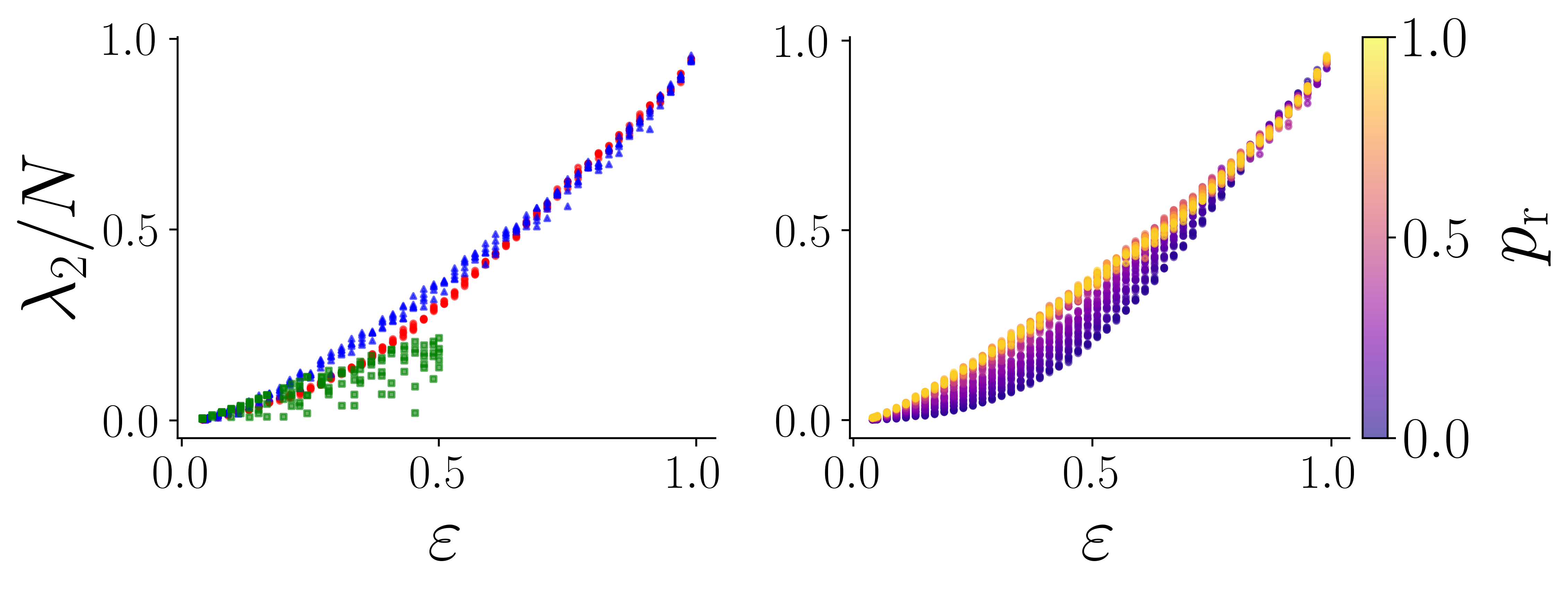}
    \caption{Relation between edge density $\epsilon$ and algebraic connectivity $\algcon$ for various coupling topologies (left: random (blue); small-world ($\rewprob=0.1$; red); scale-free (green) and for small-world topologies (right) with different rewiring probabilities $\rewprob$. Network size $N=100$. Note that the generation scheme for scale-free networks used here allows for a maximum edge density $\epsilon=0.5$.}
    \label{fig:L2vsEps}
\end{figure}
Although edge density and algebraic connectivity are often correlated~\cite{fallat2003,meng2018} --~ more edges (higher density) lead to higher algebraic connectivity, in general (cf. Fig.~\ref{fig:L2vsEps})~--, edge density only counts how many, while algebraic connectivity reflects how well vertices are connected structurally, thereby revealing hidden vulnerabilities.

\begin{figure*}[htbp]
   \centering
    \includegraphics[width=\linewidth]{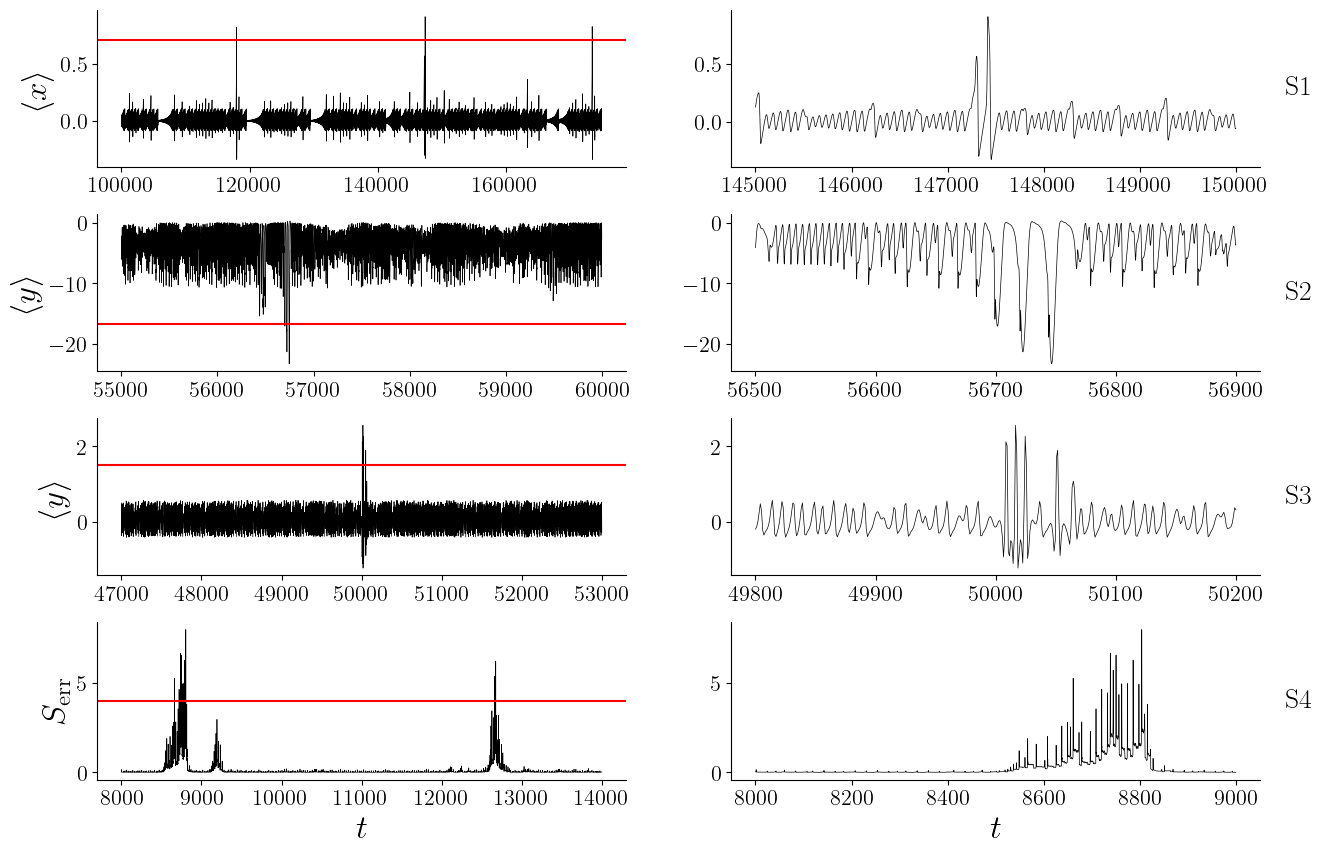}
    \caption{Left: exemplary time series of observables of the regarded systems S1~--~S4 exhibiting rare, high-amplitude events, which we consider as extreme events. Observables can be mean values of system components as in S1--S3 or the synchronization error in S4. The red horizontal line indicates the threshold $\Theta$ (Eq.~\ref{eq:threshold}). Right: excerpts centered around an extreme event.    
    }
    \label{fig:ExEvTS}
\end{figure*}
In order to identify extreme events in a time series of a system's observable (see Fig.~\ref{fig:ExEvTS} and Appendix~\ref{app:systems}), we first apply an amplitude criterion. 
A time series is considered to contain extreme events if it features high-amplitude excursions that exceed a predefined threshold
\begin{equation}
    \Theta=\text{max}(\Theta_0,\langle P\rangle +\Theta_R \sigma_P).
    \label{eq:threshold}
\end{equation}
Here, $\langle P \rangle$ and $\sigma_P$ denote mean and standard deviation of all local maxima in the time series. 
$\Theta_0$ is a constant threshold, ensuring that only high-amplitude excursions are characterized as extreme, while $\Theta_R$ ensures that excursions exceeding the threshold are highly deviating from the regular dynamics and rare~\cite{Letellier2025} (cf. Tab.~\ref{tab:integration_parameters}).
Eventually, we only assume events to be extreme if they occur non-periodically and relatively rarely. 
We discard time series with periodic~\cite{ansmann2015Chaos} occurrences of extreme events.

If an extreme event can not be identified for any of the five different realizations of a networked system, we modify (either increase or decrease, depending on the investigated system) the coupling strength $k$ by a small amount $\Delta_k$ for which we generate a new time series of the system observable using the same coupling topology and the same set of initial conditions.
This procedure is repeated until a coupling strength is identified, for which the networked system exhibits extreme events. 
Then, the increment $\Delta_k$ is decreased adaptively (by a factor of 10) to enhance the accuracy of the estimation of the coupling threshold $\kthresh$, and the analysis is repeated in a smaller interval around the previously identified coupling threshold. 
This process is repeated until a preset resolution ($\Delta_k < 10^{-4}$) is reached. 

\section{Results}
\begin{figure}[htbp]
   \centering
    \includegraphics[width=\linewidth]{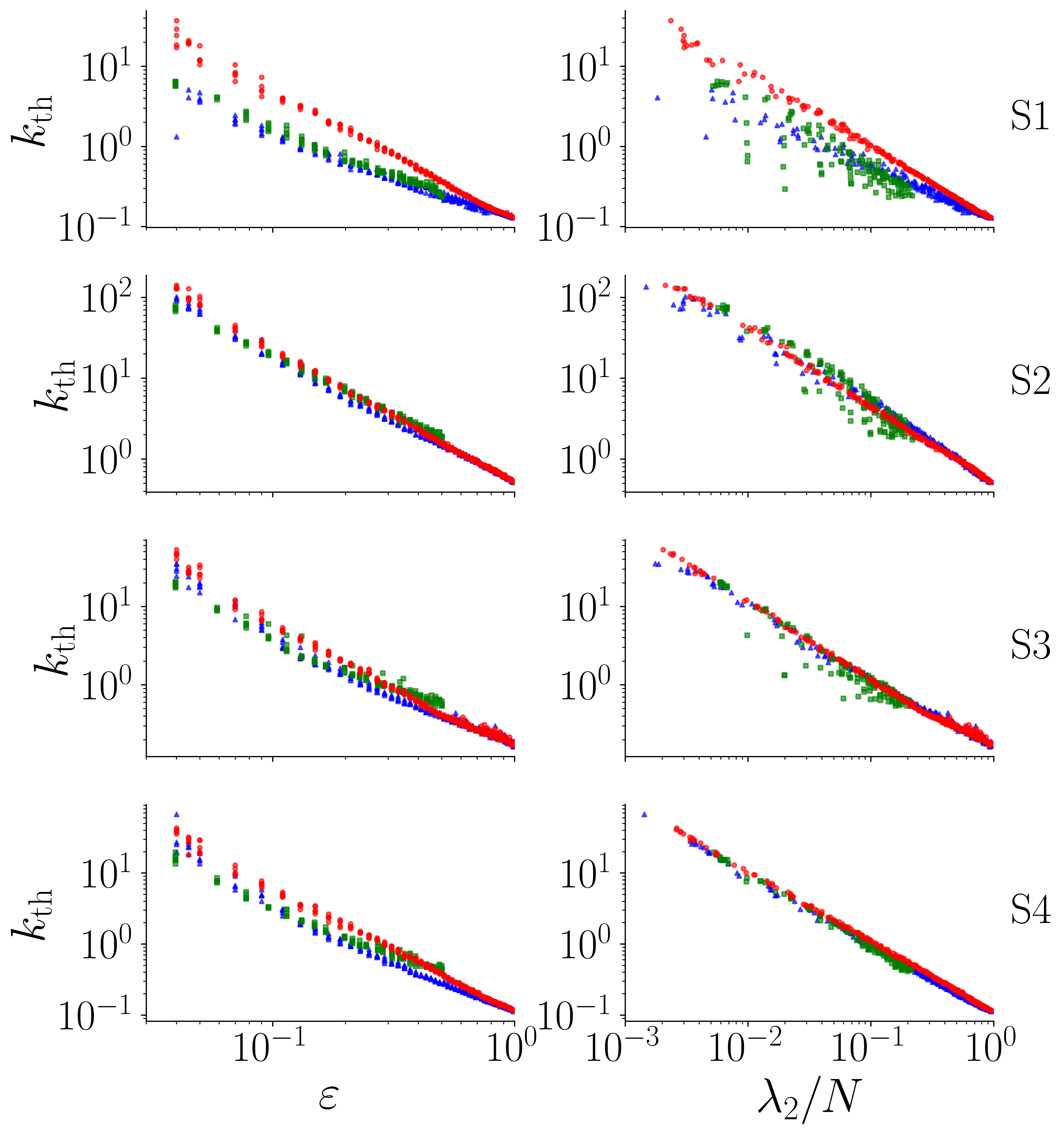}
    \caption{Relationship between coupling threshold $\kthresh$ to generate extreme events by the regarded dynamical systems S1~--~S4 and topological (edge density $\epsilon$; left) as well as spectral network properties (algebraic connectivity $\algcon$; right). Coupling topologies based on random (blue), small-world ($\rewprob = 0.25$; red), and scale-free (green) networks, each of size $N=100$. 
    }
    \label{fig:kt_eps_lambda_alltopos}
\end{figure}

In Fig.~\ref{fig:kt_eps_lambda_alltopos}, we demonstrate how the considered topological and spectral network properties relate to the coupling threshold $\kthresh$ to generate extreme events.  
Surprisingly, we observe for all networked dynamical systems and irrespective of the coupling topology, a similar qualitative relationship between edge density as well as algebraic connectivity and $\kthresh$. For the most part, this relationship can be described as a power-law in good approximation.
There are, however, some peculiarities that we briefly discuss in the following.

As can be identified from the $\epsilon-\kthresh$ relationship (Fig.~\ref{fig:kt_eps_lambda_alltopos}, left), the coupling threshold is lowest in the limiting case of globally coupled ($\epsilon \rightarrow 1$) subsystems; S1: $0.128$, S2: $\approx 0.5$, S3: $0.1 - 0.2$, and S4: $\approx 0.12$. 
This implies that the denser the network, the higher the reachability of subsystems, which requires only weak network-wide interactions (low coupling strength) for extreme events to emerge.   
The fluctuations seen for systems S2 - S4 can be traced back to different realizations of the dynamics (choice of initial conditions). 
For smaller edge densities, the coupling-topology-realization-dependent fluctuations of $\kthresh$ increase across all systems.
This is expected, since structural properties of the coupling topology, such as average clustering coefficient or average shortest path length, are sensitive to small changes in topology, especially at low edge densities. 
As a result, the value of the coupling threshold is affected.

For a wide range of edge densities, the coupling threshold  necessary to generate extreme events is higher for small-world than for the other coupling topologies. 
This is most pronounced for system S1, for which $\kthresh$
exceeds the corresponding value observed for random coupling topologies by a factor of up to four.  
For the other systems, we observe an exceedance of a factor of up to two.
We will address this issue in more detail later in this section.
For scale-free and random coupling topologies, the $\epsilon-\kthresh$ relationship is quite similar for sparser networks across systems. 
In contrast, for denser networks ($\epsilon \approx 0.3-0.5$), we observe $\kthresh$ to be higher for scale-free than for random coupling topologies. 
For systems S2 - S4, $\kthresh$ for the scale-free topologies even exceeds the respective values for the small-world topologies by about 50\%. 
This is not the case for system S1.

We now turn to the $\algcon-\kthresh$ relationship (Fig.~\ref{fig:kt_eps_lambda_alltopos}, right). 
Given the close relationship between edge density $\epsilon$ and algebraic connectivity $\algcon$ (cf. Fig.~\ref{fig:L2vsEps}), we expect a generally comparable relationship between coupling threshold $\kthresh$ and $\algcon$.
Nevertheless, there are again some, mostly system-specific peculiarities that we briefly discuss in the following.

For all systems, the power-law-like relationship between $\kthresh$ and $\algcon$ is most pronounced for small-world coupling topologies, possibly reflecting the well-known phenomenon that small worlds are easier to synchronize~\cite{nishikawa2003}.
For system S4, the power-law-like relationship holds for all coupling topologies. 
We hypothesize that the underlying mechanism (desynchronization bursts due to attractor bubbling) for a critical transition to occur overrides the possible impact of the coupling topology.
For system S3, which approaches complete synchronization as the coupling strength increases, the power-law-like relationship is indefinite for random and even more undefined for scale-free coupling topologies.
We observe even stronger deviations for systems S2 and S1, although deviations decrease for larger network sizes (see Appendix~\ref{app:N}, Fig.~\ref{fig:kth-N}).
We hypothesize that deviations may be either due to the comparably larger control parameter mismatches (affecting the eigendynamics of the respective subsystems) or due to the underlying mechanism (interior crisis) for a critical transition to occur in these systems, or both.

\begin{figure}[htbp]
   \centering
    \includegraphics[width=\linewidth]{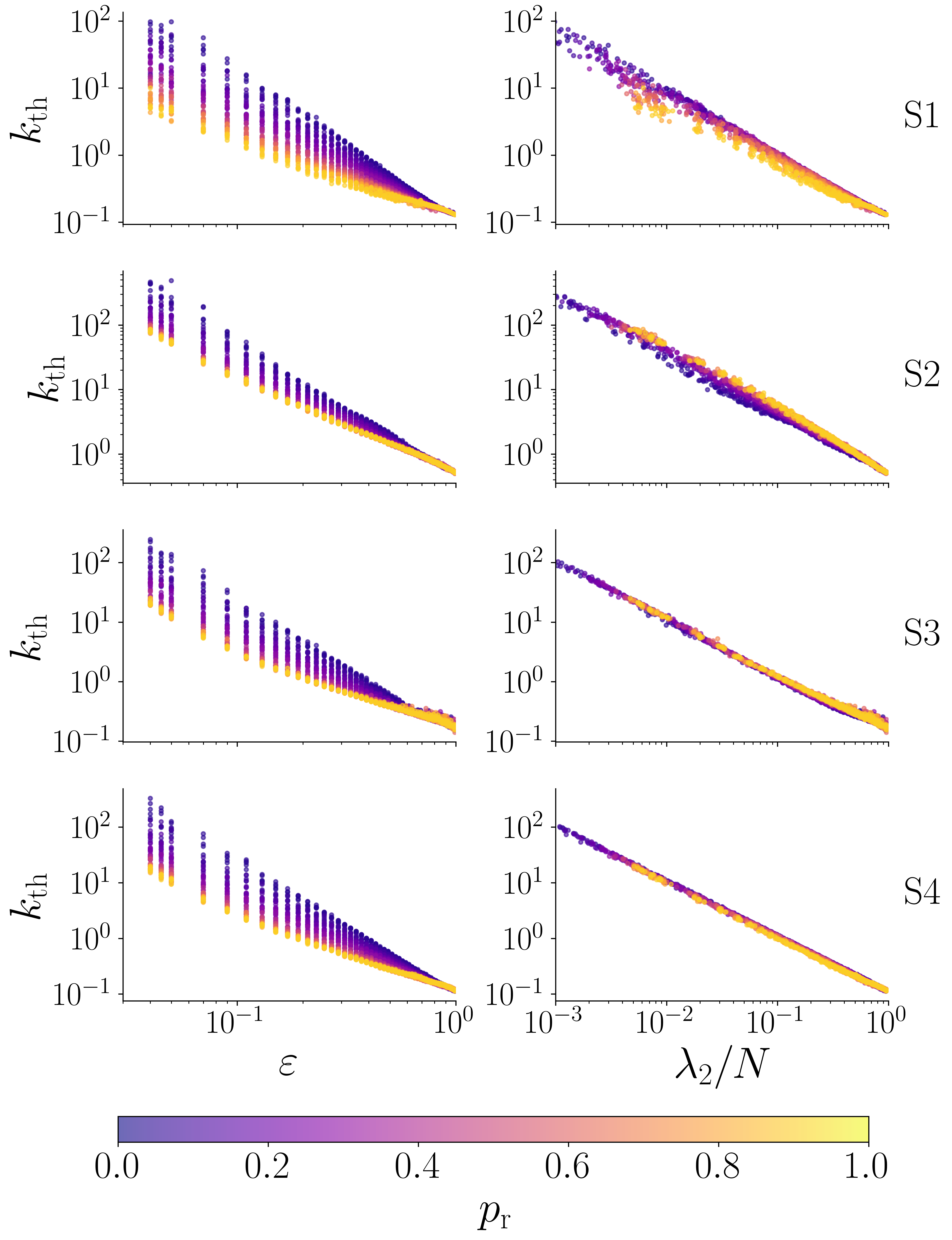}
    \caption{Same as Fig.~\ref{fig:kt_eps_lambda_alltopos}, but for coupling topologies based on small-world networks with different rewiring probabilities $\rewprob$.}
    \label{fig:kt_eps_lambda_SW}
\end{figure}

Despite these differences the observed relationship between coupling threshold $\kthresh$ and edge density $\epsilon$, resp., algebraic connectivity $\algcon$ is similar across the different systems and across different coupling topologies. 
Nevertheless, the small-world coupling topology appears to play a special role, indicating that shortcuts in the coupling topology may impact the coupling threshold necessary to generate extreme events.
Figure~\ref{fig:kt_eps_lambda_SW} summarizes our findings obtained from varying the rewiring probability $\rewprob$.
We again observe a power-law-like relationship between edge density $\epsilon$ and coupling threshold $\kthresh$ for all systems and all rewiring probabilities (Fig.~\ref{fig:kt_eps_lambda_SW}, left), however, with different slopes. 
For a given edge density, a smaller $\rewprob$ tends to result in a higher $\kthresh$. 
Particularly for very sparse small-world coupling topologies, the number of shortcuts has, in general, a significant impact on $\kthresh$ (data not shown). As an example, we find for system S1 with $\epsilon =0.04$ a coupling threshold $\kthresh \approx 5$ for $\rewprob=0.9$ but $\kthresh \approx 100$
on a small-world coupling topology with only a few shortcuts ($\rewprob=0.05$).
Even for constant edge densities, the coupling threshold can vary across orders of magnitude for different rewiring probabilities.
Interestingly, shortcuts in the coupling topology appear to have only a very minor influence on the relationship between algebraic connectivity $\algcon$ and coupling threshold $\kthresh$ (Fig.~\ref{fig:kt_eps_lambda_SW}, right). 
Deviations from a power-law-like relationship are mostly system-specific, with more pronounced deviations seen for systems S1 and S2.
As before, we hypothesize that deviations may be either due to the comparably larger control parameter mismatches or due to the underlying mechanism (interior crisis) for a critical transition to occur in these systems, or both.

\begin{figure*}[htbp]
   \centering    \includegraphics[width=\linewidth]{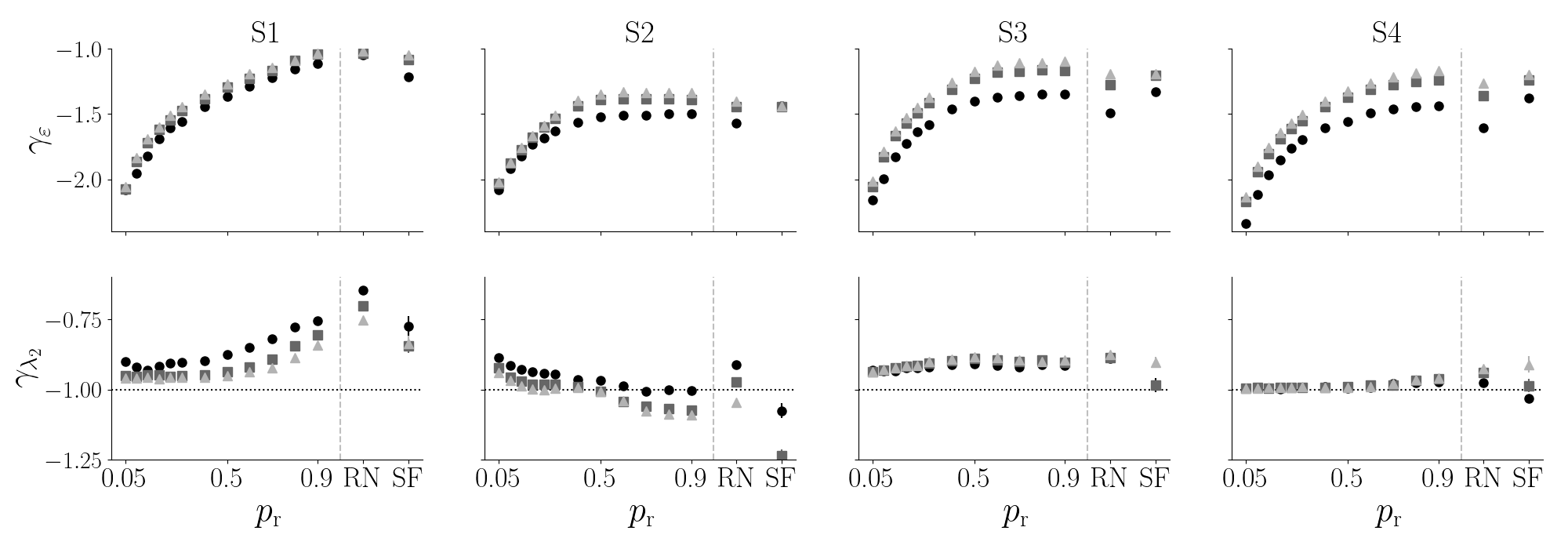}
    \caption{Power law exponents $\gameps$ and $\gamlam$ extracted from the data shown in Figs.~\ref{fig:kt_eps_lambda_alltopos}~and~\ref{fig:kt_eps_lambda_SW}. 
    Different symbols denote different coupling topologies: random (RN)---triangles, scale-free (SF)---squares, small-world---circles.
    Numbers on the x-axis indicate the rewiring probability $\rewprob$ used to generate small-world coupling topologies. Different markers denote networks sizes: $N=100$~--~circles, $N=200$~--~squares, $N=300$~--~triangles. Error bars are smaller than symbol size.}
    \label{fig:exponents}
\end{figure*}
Eventually, we quantify the relationship between  coupling threshold $\kthresh$ and edge density $\epsilon$, resp. algebraic connectivity $\algcon$ and determine the corresponding power law exponents $\gameps$ and $\gamlam$ by fitting a linear function to the logarithmically transformed data.
Figure~\ref{fig:exponents} summarizes our findings for all model systems.
We observe the largest magnitude of $\gameps$ for very sparse small-world coupling topologies ($\rewprob =0.05$), varying between $\gameps\approx-2$ for system S1 and $\gameps\approx-2.5$ for system S4.   
For these systems, $\gameps$ steadily increases with increasing the rewiring probability $\rewprob$, while for systems S2 and S3  $\gameps$ increases until $\rewprob\approx 0.5$ and then saturates. 
For $\rewprob = 0.9$, the exponent varies across systems between $\gameps \approx -1.1$ for system S1 and $\gameps \approx -1.5$ for system S2. 
The exponents seen for the random and scale-free coupling topologies are, in general, similar to those of the small-world topologies, with $\rewprob = 0.9$.
For systems S3 and S4, however, the magnitude of $\gameps$ is slightly higher for the random coupling topologies. 
The exponents differ for different network sizes $N$. 
For system S1, the difference is only weakly pronounced.
For the other systems, $\gameps$ converges as $N$ increases.

As already expected from the findings shown in Figs.~\ref{fig:kt_eps_lambda_alltopos}~and~\ref{fig:kt_eps_lambda_SW}, we observe $\gamlam \approx -1$ for systems S3 and S4, mostly independent of the coupling topology. 
Up to some fluctuations in the percent range (that vanish as $N$ increases; cf. Appendix~\ref{app:N}, Fig.~\ref{fig:kth-N}), the data reflect a relationship $\kthresh \propto \algcon^{-1}$, consistent with the assumption of the Master Stability Formalism\cite{pecora1998,sun2009} $k\algcon = \rm{const}$, which guarantees the stability of the synchronized/near-synchronous state.
For systems S1 and S2, the data do not show a power law with similar exponents, independent of the coupling topology. 
For small-world coupling topologies, we observe the magnitude of $\gamlam$ to decrease as $\rewprob$ increases for system S1, while to opposite is true for system S2. 
In general, $\gamlam$ appears to converge as $N$ increases.
We conjecture that the varying exponents seen for systems S1 and S2 and possibly also for S3 result from the different mechanisms leading to extreme events, and aspects of the subsystems' eigendynamics, such as parameter mismatch, may still play a role.

\section{Discussion}
Improving our understanding of how extreme events emerge in networked dynamical systems through model studies is often hindered by the need to carefully select control parameter settings for both vertex dynamics and coupling topology.
Here, we asked whether there exists a dynamics- and coupling-topology-dependent threshold for the coupling strength $\kthresh$ that leads to the emergence of extreme events in the collective dynamics of such systems.
To this end, we investigated various networked systems in which extreme events occur via distinct routes, including an interior crisis (with and without intermittency) and a bubbling transition.
Across these systems, for different paradigmatic coupling topologies and edge densities, we determined $\kthresh$ by a brute-force search and found robust power-law-like relationships between $\kthresh$ and the network's edge density $\epsilon$ as well as between $\kthresh$ and the algebraic connectivity $\algcon$, i. e., the second smallest eigenvalue of the network Laplacian.
These relationships emerge largely independently of the specific route by which extreme events are generated, which indicates that topological and spectral properties of the coupling structure play a dominant role in setting the coupling threshold for extreme-event generation.

For systems in which extreme events are associated with the (in-)stability of a synchronized or near-synchronized state, we obtained $\kthresh \propto \algcon^{-1}$ with good accuracy.
This scaling is consistent with expectations from the Master Stability Formalism and related linear stability analyses of synchronization, where products of coupling strength and Laplacian eigenvalues determine stability~\cite{wu1996,Nazerian2024,CubillosCornejo2025}.
In contrast, for systems in which extreme events arise from crisis-induced transition, we still observe approximate power-law relationships, but with system-dependent exponents and more pronounced deviations at small $\algcon$.
Our inconclusive observations when considering the largest Laplacian eigenvalue $\lambda_{N}$~\cite{Castellano2017,Faci2023} may indicate that the onset of extreme events is more directly constrained by the weakest transverse mode, captured by $\algcon$, than by the overall spectral radius.

Although coupling topology and its spectral properties appear to be the primary determinants for $\kthresh$, our results also reveal systematic differences across the investigated systems, for example in the values of the power-law exponents.
These differences suggest that aspects of the intrinsic dynamics at each vertex, such as parameter mismatches or the precise transition scenario underlying the generation of extreme events, modulate the scaling laws rather than being entirely negligible.
Clarifying this interplay between topology and local dynamics, for example by varying parameter mismatches or by analyzing reduced normal forms, is a natural direction for future work.

Our study is restricted to linear diffusive coupling on pairwise networks.
An important next step is to test whether similar scaling relationships between $\kthresh$ and spectral/topological properties persist for more general coupling architectures, such as higher-order and multilayer networks~\cite{battiston2026}, or for qualitatively different collective phenomena, including explosive synchronization~\cite{Boccaletti2016} or the onset of chimera states~\cite{Majhi2019}.
Such extensions would help to delineate which aspects of the observed scaling are generic to the occurrence of extreme events in networked dynamical systems and which are specific to the coupling form and generating mechanism.

Our results also raise interesting questions for real-world networks in which the topology can be modified more easily than the local dynamics.
If relationships of the form $\kthresh \propto \epsilon^{\gamma_\epsilon}$ or $\kthresh \propto \algcon^{\gamma_{\algcon}}$ remain qualitatively valid in more realistic settings, then relatively coarse structural interventions --~such as changing edge density by adding or removing links, or rewiring connections to adjust $\algcon$~-- could shift the onset of extreme events by orders of magnitude in coupling strength, even when the vertex dynamics is fixed.
In infrastructures, this would mean that choices about redundancy and sparsification (for example, whether to add long-range shortcuts or to concentrate links on hubs) do not just affect classical robustness measures, but also the range of operating conditions under which rare, catastrophic excursions can occur.
In neuronal or ecological networks, a similar logic would suggest that plasticity or adaptation processes that reshape connectedness of network constituents could strongly modulate the propensity for extreme events, even in the absence of changes in single-unit dynamics.
In addition, in various systems the occurrence of self-organized criticality is hypothesized to occur prior to the formation of extreme events. 
This relationship between network connectedness and the probability of the emergence of extreme events could shed some light on the mechanism of emerging self-organized criticality by reordering connections in the network~\cite{Meisel2009,Sugimoto2025} to achieve a critical state in which extreme events can occur.	 
Exploring how such scaling laws manifest in data-driven network models, and how they interact with constraints specific to each domain, is an important step toward translating our findings into concrete design and control strategies.
Our scaling relationships may also provide information about the rate of occurrence and about the amplitude distribution of extreme events~\cite{Sornette2012,Mishra2018,Das2024} once the system is above the dynamics- and coupling-topology-dependent threshold for the coupling strength. 
Understanding whether denser or more algebraically connected networks produce more frequent or more intense extreme events could have important practical implications~\cite{Sachs2012}.

Taken together, our findings provide quantitative scaling relationships that can guide future model studies in selecting control parameter settings that give rise to extreme events in networked dynamical systems.
By relati,ng the coupling threshold directly to easily accessible network characteristics, they allow one to obtain order-of-magnitude estimates of $\kthresh$ without exhaustive parameter scans, and thereby substantially reduce the computational effort required for exploration.
If similar relationships can be established and calibrated for specific real-world systems, they may eventually inform the design or modification of networks~\cite{Sahneh2012,Wang2013} to mitigate the risk of extreme events or to shift their onset to less critical parameter regimes.

\begin{acknowledgments}
The authors acknowledges fruitful discussions with Manuel Adams, Max Potratzki, and Peter Ashwin. 
KL acknowledges support from the Deutsche Forschungsgemeinschaft.
\end{acknowledgments}


\section*{AUTHOR DECLARATIONS}

\subsection*{Conflict of Interest}
The authors have no conflicts of interest to disclose.

\subsection*{Data availability}
The data that support the findings of this study are available from the corresponding author upon reasonable request.


\newpage
\appendix
\section{Model systems}
\label{app:systems}
\subsection*{S1: Networks of coupled FitzHugh-Nagumo oscillators}
Coupled oscillators of the FitzHugh-Nagumo type are a paradigmatic model for neural dynamics and can exhibit rich dynamical behavior, including extreme events~\cite{ansmann2013PhysRevE,karnatak2014PhysRevE,Saha2017,Broehl2020,Gerster2020,CubillosCornejo2025,Hariharan2026,Hariharan2026b} and self-induced switching between multiple space-time patterns~\cite{Ansmann2016}.
We here consider $N$ diffusively coupled FitzHugh–Nagumo oscillators $\left(i\in \left\{1, \ldots, N \right\}\right)$ whose $i$th oscillator is described by
\begin{align*}
    \dot{x}_i &= x_i(a-x_i)(x_i-1)-y_i +\frac{k}{N-1}\sum_{j=1}^N A_{ij}(x_j-x_i),\\
    \dot{y}_i &= b_ix_i-cy_i.
\end{align*}
Here, $a=-0.02651$, $b_i = 0.006+\frac{i-1}{N-1}\times 0.008$ and $c = 0.02$ are fixed internal control parameter~\cite{ansmann2013PhysRevE,karnatak2014PhysRevE}.
We chose initial conditions for the $x$ variable from ${\cal N}(0.15, 0.6)$ and for the $y$ variable from ${\cal N}(0.15, 0.1)$. 
The sample average $\langle x\rangle =\frac{1}{N}\sum_{i=1}^{N} x_i$ served as the observable for this system. 

\subsection*{S2: Networks of coupled memristive Hindmarsh-Rose neurons}
The Hindmarsh–Rose model of neuronal activity~\cite{Hindmarsh1984} mimics the spiking-bursting behavior of the membrane potential observed in experiments made with a single neuron.
The model modification consider here~\cite{vijay2024EurPhysJ} employs an active flux-controlled memristor~\cite{Chua2003} that takes into account electromagnetic induction effects between coupled neurons.
Upon varying the membrane current $I$, the coupled neuron mode transits from bounded chaotic spiking oscillations to superextreme spiking oscillations~\cite{Vijay2023}.
We here consider $N$ diffusively coupled Hindmarsh-Rose neurons $\left(i\in \left\{1, \ldots, N \right\}\right)$ whose $i$th neuron is described by
\begin{align*}
    \dot{x}_i &= y_i +bx_i^2 -a x_i^3 +I_i +gz_ix_i +\frac{k}{N-1}\sum_{j=1}^N A_{ij}(x_j-x_i)\\
    \dot{y}_i &= c-dx_i^2 -y_i\\
    \dot{z}_i &= x_i.
\end{align*}
Here, $a=1.0$, $b=3.2$, $c=1.0$, $d=5.0$, $g=0.9$, and $I_i = 1.0+\frac{i-1}{N-1}\times 0.01$ are fixed internal control parameter~\cite{vijay2024EurPhysJ}. 
We chose initial conditions for the $x$ variable from ${\cal N}(0.5, 0.5)$, for the $y$ variable from ${\cal N}(0.2, 0.5)$, and for the $z$ variable from ${\cal N}(0.1, 0.5)$.
The sample average $\langle y\rangle =\frac{1}{N}\sum_{i=1}^{N} y_i$ served as the observable for this system. 

\subsection*{S3: Networks of coupled periodically forced Li\'enard-type oscillators}
Li\'enard systems constitute a general class of two-dimensional autonomous systems and can be written as $\ddot{x}+f(x)x+g(x)=0$.
Nonlinear oscillators of Li\'enard-type have been used to model numerous physical phenomena ranging from atmospheric physics, condensed matter and nonlinear optics to electronics, plasma physics, biophysics, and evolutionary biology.
Li\'enard-type oscillators with an external sinusoidal forcing can generate extreme events for a suitable choice of parameters~\cite{kingston2017PhysRevE}.
We here consider $N$ coupled such oscillators $\left(i\in \left\{1, \ldots, N \right\}\right)$ whose $i$th oscillator is described by
\begin{align*}
    \dot{x}_i &= y_i,\\
    \dot{y}_i &= -\alpha x_iy_i -\gamma x_i -\beta x_i^3 +F\sin(\omega t) +\frac{k}{N-1}\sum_{j=1}^N A_{ij}(y_j-y_i).
\end{align*}
Here, $\alpha = 0.45$, $\beta=0.5$, $\gamma=-0.5$, and $F=0.2$ are fixed internal control parameter, and with $\omega=0.7325$ the system is capable of generating extreme events via an interior crisis~\cite{kingston2017PhysRevE}.
We chose initial conditions for $x$ and $y$ randomly from the interval $[-2,2]$.
The sample average $\langle y\rangle =\frac{1}{N}\sum_{i=1}^{N} y_i$ served as the observable for this system. 

\subsection*{S4: Networks of coupled R\"ossler oscillators}
The R\"ossler oscillator was proposed by Otto E. R\"ossler~\cite{Roessler1976} as a simplified Lorenz oscillator~\cite{Lorenz1963} that describes convection in the earth’s atmosphere. 
The oscillator is capable of generating a wide range of dynamical behaviors, including chaos, and a network of nearly identical R\"ossler oscillators gives rise to desynchronization events, known as bubbling~\cite{tirabassi2025Arxiv}.
We here consider $N$ coupled such oscillators $\left(i\in \left\{1, \ldots, N \right\}\right)$ whose $i$th oscillator is described by
\begin{align*}
    \dot{x}_i &= -y_i -z_i +\frac{k}{N-1} \sum_{j=1}^N A_{ij}(x_j-x_i),\\
    \dot{y}_i &= x_i +a_iy_i +\frac{k}{N-1}\sum_{j=1}^N A_{ij}(y_j-y_i),\\
    \dot{z}_i &= b_i + x_i(z_i-c) +\frac{k}{N-1}\sum_{j=1}^N A_{ij}(z_j-z_i).
\end{align*}
Here, $a_i = b_i = 0.199+\frac{i-1}{N-1}\times 0.002$ and $c_i = 6.965+\frac{i-1}{N-1}\times 0.07$ are fixed internal control parameter~\cite{tirabassi2025Arxiv}.
We chose initial conditions for $x$, $y$, and $z$ randomly from the interval $[0,1]$.
The synchronization error
\begin{align*}
S_{\rm err}(t) = \sum_{i=1}^{N} \sqrt{ \left( x_i- \langle x \rangle \right)^2 + \left( y_i-\langle y \rangle \right)^2 + \left( z_i- \langle z \rangle \right)^2 } / N,
\end{align*}
where $\langle \cdot \rangle$ denotes the sample average, served as the observable for this system.

\subsection*{Generating time series}
All systems were integrated using the Dormand-Prince method~\cite{ansmann2018Chaos} with an integration step size $\tau$.  
We generated time series of system observables of length $T$ after discarding $T_{\rm trans}$ transients.
Table~\ref{tab:integration_parameters} summarizes the corresponding parameters for each system. 
\begin{table}[h]
    \centering
    \begin{tabular}{c|ccc|cc}
    system     & $T$&$\tau$ &$T_{\rm trans}$&$\Theta_0$&$\Theta_R$ \\
    \hline
    S1 & $ 2\cdot 10^5 $ & $1.0$ & $0.5\cdot 10^5$ & $0.6$ &$8$\\
		S2 & $ 1\cdot 10^5 $ & $0.1$ & $0.2\cdot 10^5$ & $15$  &$4$\\
    S3 & $ 1\cdot 10^5 $ & $1.0$ & $0.2\cdot 10^5$ & $1.5$ &$8$\\
    S4 & $ 1\cdot 10^4 $ & $0.1$ & $0.5\cdot 10^4$ & $1$   &$8$\\
    \end{tabular}
    \caption{Integration parameters for the generation of time series for each investigated system. 
		$\Theta_0$ and $\Theta_R$ are the system-specific control parameters used to identify extreme events.}
    \label{tab:integration_parameters}
\end{table}

\section{Impact of network size}
\label{app:N}
\begin{figure}[h]
   \centering \includegraphics[width=\linewidth]{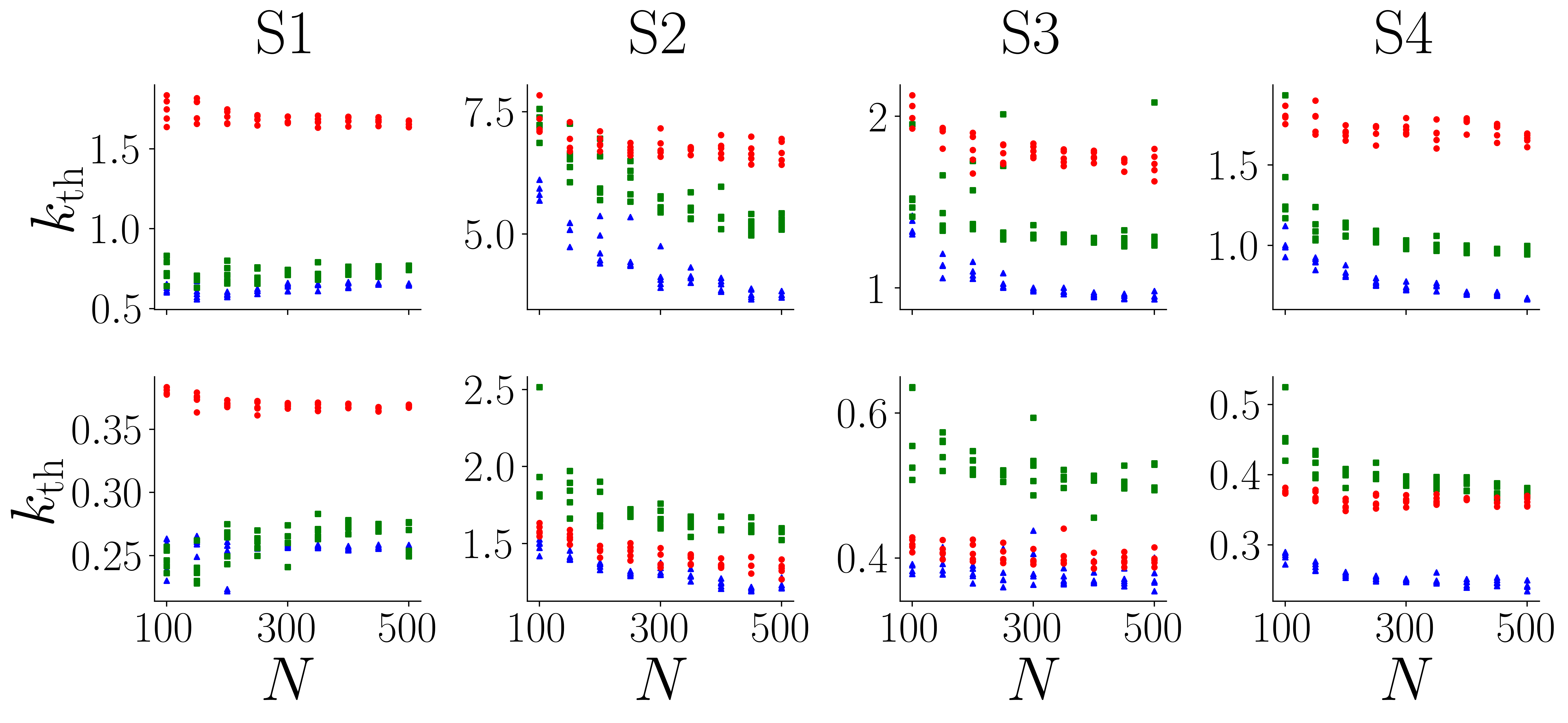}
    \caption{Estimated coupling threshold $\kthresh$ for the networked systems S1~--~S4 with different sizes $N$ and for edge densities $\epsilon = 0.2$ (upper panel) and $\epsilon=0.5$ (bottom panel). Coupling topologies based on random (blue), small-world ($\rewprob=0.25$; red) and scale-free (green) networks.}
    \label{fig:kth-N}
\end{figure}

Fig.~\ref{fig:kth-N} shows how the coupling threshold $\kthresh$ varies for different realizations of the networked systems S1 -- S4 of different sizes $N$.
 
%

\end{document}